\documentclass[twocolumn,showpacs,english,prb,preprintnumbers,amsmath,amssymb,floatfix,eqsecnum]{revtex4}
\usepackage[T1]{fontenc}
\usepackage[latin9]{inputenc}
\usepackage{babel}
\usepackage{graphicx}
\usepackage{epsfig,psfrag}
\usepackage{dcolumn}
\usepackage{bm}
\usepackage{color}
\usepackage{esint}
\begin{document}
\title{Anomalous Josephson current, incipient time-reversal 
symmetry breaking, and Majorana bound states in interacting multi-level dots}
\author{Aldo Brunetti,$^1$ Alex Zazunov,$^1$ Arijit Kundu,$^2$
 and Reinhold Egger$^1$}
\affiliation{$^1$~Institut f\"ur Theoretische Physik,
Heinrich-Heine-Universit\"at, D-40225  D\"usseldorf, Germany\\
$^2$~Department of Physics, Indiana University, 727 East Third Street, Bloomington, IN 47405-7105 USA}
\date{\today}
\begin{abstract} 
We study the combined effects of spin-orbit interaction, magnetic field,
and Coulomb charging on the Josephson current-phase relation,
$I(\varphi)$, for a multi-level quantum dot tunnel-contacted by
two conventional $s$-wave superconductors with phase 
difference $\varphi$.  A general model is formulated and analyzed in
the cotunneling regime (weak tunnel coupling)
and in the deep subgap limit, fully taking into account interaction effects.  
We determine the conditions for observing
a finite anomalous supercurrent $I_a=I(\varphi=0)$.  For a 
two-level dot with spin-orbit coupling and arbitrarily weak Zeeman field $B$, 
we find the onset behavior $I_a\propto {\rm sgn}(B)$
in the presence of interactions,
suggesting the incipient spontaneous breakdown of time-reversal symmetry.
We also provide conditions for realizing spatially separated 
(but topologically unprotected) Majorana bound states in a double dot
variant of this system. Here Majoranas are predicted to leave
a clear signature in the $2\pi$-periodic current-phase relation.
\end{abstract}
\maketitle

\section{Introduction}\label{sec1}

Studies of the current-phase relation (CPR) in a Josephson junction, 
where a weak link connects two superconductors with 
phase difference $\varphi$, 
have provided ever new surprises over the past fifty years.\cite{jj}  
Nowadays, Josephson junctions showing novel and 
rich behavior can be formed by sandwiching a nanoscale
conductor -- collectively referred to as 'quantum dot' below, e.g., 
a semiconductor dot or nanowire, or 
a single molecule -- between two superconductors.\cite{silvano,alfredo} 
The interest in such nanoscale hybrid devices 
has sharply increased recently due to technological
advances, allowing to fabricate and manipulate well-characterized
setups and raising the hope for new applications, 
as well as by the prospect of realizing Majorana fermions.  
To mention just a few key experiments, gate-tunable supercurrents through
the two-dimensional electron gas in semiconductors 
have been demonstrated,\cite{exp1,exp2,exp3,exp4,exp5}
the CPR of superconducting atomic point contacts has been measured using a loop 
geometry,\cite{rocca} and the direct spectroscopy of Andreev bound states 
in carbon nanotube devices was reported.\cite{pillet} 
The phenomena studied below will be particularly pronounced
for strong spin-orbit coupling (SOC) in the nanoscale conductor.
Note that strong SOC is naturally present in InAs or InSb,\cite{winkler,doh,dam,nbi,lee1,chang,lee2,deacon,kanai,nilsson}
and in self-assembled SiGe quantum dots.\cite{katsaros}
SOC is often responsible for nontrivial topological properties 
and the emergence of Majorana fermions in very similar 
settings.\cite{hasan,qi,carlo,alicea,karsten,lutchyn,oreg}
Majoranas have attracted wide attention after recent experiments reported first 
transport signatures such as those expected for Majorana 
fermions.\cite{mourik,rokhinson,das,deng}

In this paper, we study a general model for 
the equilibrium Josephson current through a multi-level quantum dot 
tunnel-contacted by two conventional $s$-wave BCS superconductors with
phase difference $\varphi$ and superconducting gap $\Delta$.  
Our dot Hamiltonian $H_d$, see Eq.~\eqref{hd} below, takes into account
arbitrary SOC terms, magnetic (orbital and Zeeman) field effects, and
Coulomb charging interactions.  Moreover, the tunnel contacts are
described by a general tunneling Hamiltonian $H_t$, see Eq.~\eqref{ht} 
below, allowing for inter-orbital phase shifts and asymmetric contacts.
Our analysis is mostly devoted to two complementary regimes where 
analytical progress is possible, namely the cotunneling regime,
realized for weak tunneling, and  the 
deep subgap regime ('atomic limit'), where $\Delta$ represents
the largest energy scale.  We explore in detail the ground-state 
Josephson CPR, $I(\varphi)$, which can reveal two particularly interesting  
phenomena in such a setting, namely the anomalous Josephson effect
 and Majorana bound states (MBSs). 

The anomalous Josephson effect is characterized by a
finite supercurrent flowing at zero phase difference,
$I_a\equiv I(\varphi=0)\ne 0$. Comparing to the conventional Josephson
relation, $I(\varphi)=I_c \sin\varphi$ with critical
current $I_c$,  this is equivalent to a $\varphi_0$ phase shift, i.e.,
$I_a=I_c\sin\varphi_0$. Junctions with $I_a\ne 0$ are thus commonly
referred to as '$\varphi_0$-junctions', where  SOC is typically a 
crucial ingredient.  The Josephson CPR for quantum dots with SOC has been
studied in many theoretical works,\cite{bezuglyi,krive1,krive2,feigelman,prbSO,beri,feinberg,buzdin,prlSO,liu,denis2,nazarov}
and the conditions for $\varphi_0$-junction behavior have been clarified in the 
noninteracting case.\cite{krive1,krive2,feinberg,buzdin,prlSO} 
In contrast to the widely known $0$- and $\pi$-junctions,\cite{jj} 
where $\varphi_0=0$ and $\varphi_0=\pi$, respectively,
a general $\varphi_0$-junction can have direction-dependent
critical currents,\cite{feinberg,denis2} i.e., 
$I_{c1}={\rm max}[I(\varphi)]$ and $I_{c2}={\rm max}[-I(\varphi)]$ are
different.  The $\varphi_0$-junction can thus 
act as a phase battery\cite{goldobin}
or as superconducting rectifier,\cite{feinberg,denis2} promising novel
device applications. 
While it is well-established\cite{goldobin,nazarov1,kschan,brydon,linder} and also
experimentally observed\cite{sickinger} that spin-active interfaces,
e.g., for a ferromagnetic 'dot' region, allow one to realize a
$\varphi_0$-junction,  we here focus on semiconducting or 
molecular systems with spin-conserving and spin-independent 
interfaces, where $\varphi_0$-junction behavior is quite nontrivial. 
$\varphi_0$-junctions were also predicted but never observe
in unconventional superconductors.\cite{unconv1,unconv2,unconv3,unconv4,trs1}

So far, the necessary conditions for anomalous supercurrents have only
been determined for noninteracting dots, where one needs
finite SOC and a suitably oriented magnetic field. In addition, 
asymmetric tunnel contacts with non-commuting hybridization
matrices, $\Gamma^{(L)}\ne 
\Gamma^{(R)}$, are required. This imposes a chirality condition which
is necessary to have $I_a\ne 0$, see Ref.~\onlinecite{prlSO}
and Sec.~\ref{sec3} below.  We find that 
 the Coulomb charging energy $E_c$ does not change these necessary
conditions, but it can be responsible
for a dramatic enhancement of the anomalous supercurrent.
The most interesting enhancement is related to an interaction-induced 
behavior 
with $I_a\propto {\rm sgn}(B)$
for arbitrarily weak time-reversal symmetry (TRS) breaking field $B$.
Such a behavior suggests that TRS is spontaneously broken. 
However, thermal fluctuations can suppress $I_a$, and we therefore
interpret this enhancement of $I_a$ 
compared to the usual noninteracting behavior,\cite{prlSO} 
 $I_a(B\to 0)\propto B$, as 'incipient' spontaneously broken TRS.
This effect generally happens whenever two $B=0$  
Kramer's partner states contribute with opposite sign to $I_a$.
A small magnetic field then lifts the degeneracy, while the Coulomb
interactions create a gap and effectively project away the higher energy state.
As a consequence, interaction-induced enhancement is especially
pronounced for small $B$ and if $E_c$ exceeds all 
other energy scales of interest.  Concrete parameter regimes where this
effect occurs will be discussed in Sec.~\ref{sec4}.
We mention in passing that spontaneously broken TRS was also reported in 
a recent mean-field study\cite{trs1} for a
single-level Anderson dot between a two-band ($s_\pm$) and a single-band 
($s$-wave) superconductor. However, this effect can be traced back to phase frustration\cite{trs1} and strongly differs from our scenario. 
Technically related works have also studied the supercurrent 
in the cotunneling regime for dots coupled to a 
local phonon mode\cite{novotny} and to a two-level system.\cite{schulz}
Other studies of the Josephson effect for interacting double dots have either disregarded SOC\cite{loss,aguado,meden} or did not address the phenomena investigated here.\cite{droste}

Besides analyzing the anomalous supercurrent, in Sec.~\ref{sec5} we also
 address the possibility of MBS formation in an interacting double dot 
with SOC and Zeeman field.  
The double dot is contained as special case in our general multi-level
Hamiltonian, and our theory is directly applicable to such a 
two-orbital case with well separated orbitals. 
Majorana fermions are emergent quasi-particles that equal their own antiparticle. They are of much interest  in the context of topological quantum computation.\cite{hasan,qi,carlo,alicea,karsten}
When our 'dot' region corresponds to a semiconductor nanowire,
one effectively can realize Kitaev's chain model which
(in the right parameter regime) allows for a pair of topologically
protected MBSs localized near the nanowire ends.\cite{lutchyn,oreg}
('Topological protection' implies that small parameter changes not closing a bulk gap cannot remove the MBSs.) 
As discussed by Lejinse and Flensberg,\cite{flensberg} 
see also Refs.~\onlinecite{wright,fulga}, a simpler variant, albeit 
with topologically unprotected Majorana fermions, can be 
realized for two Coulomb-blockaded single-level dots 
coupled to a superconductor. Similarly, in our setting a 
pair of spatially separated MBSs can also be realized.  
Remarkably, these Majoranas could be detected through the
highly unusual features in the $2\pi$-periodic CPR described below.

The structure of the remainder of this article is as follows.
In Sec.~\ref{sec2}, a general model for the S-Dot-S hybrid structure is
introduced.  We allow for arbitrary single-particle Hamiltonians
in the dot region, and take into account Coulomb charging effects.
Integrating out the noninteracting fermions in the superconducting 
electrodes, we arrive at an effective partition function expressed
in terms of dot variables only, which then allows to extract the 
Josephson CPR by a phase derivative.  
For concrete results, we employ a generic two-orbital dot with
a Zeeman field and (Rashba or Dresselhaus) SOC. 
In Sec.~\ref{sec3}, we discuss the two approaches used in this work.
First, we study the cotunneling regime by perturbation theory in the tunnel
couplings.  The general ground-state CPR is derived, see Eq.~\eqref{cotun},
with $I_a$ expressed in terms of matrices $J$ and $Q$, 
see Eq.~\eqref{Ica}.   $J$ depends only on single-particle 
quantities and imposes necessary conditions for $I_a\ne 0$, while  $Q$ 
encapsulates interaction effects.
As second approach, we study the 'atomic limit', $\Delta\to \infty$,
where the proximity effect of the superconducting leads is contained in an
effective dot Hamiltonian.
In Sec.~\ref{sec4}, we address the anomalous Josephson effect for
a two-level dot,  and in Sec.~\ref{sec5}, we show that a 
pair of spatially separated MBSs
emerges for suitably chosen parameters in a double dot device.   
Finally, we offer some
concluding remarks in Sec.~\ref{sec6}.  We often 
use units with $\hbar=e=k_B=1$.

\section{Model and effective partition function}\label{sec2}

\subsection{General model}

We study a general model describing the Josephson effect in 
a large variety of interacting nanostructures,
where a central region ('dot') is 
tunnel-coupled to two conventional $s$-wave superconducting leads,
$H=H_d+H_t+H_l$.  Following standard arguments,\cite{nazarovbook} we 
take into account  Coulomb interactions, SOC, and
magnetic field effects only on the dot, 
but not in the bulk electrodes nor in the tunnel contact.
For $M$ relevant (spin-degenerate) electronic orbitals in the
central dot region,  the dot Hamiltonian is taken in the form
\begin{equation}\label{hd}
H_d= \sum_{n\sigma,n'\sigma'} d_{n\sigma}^\dagger h^{}_{n\sigma,n'\sigma'}
d_{n'\sigma'}^{} + E_c (\hat N-n_g)^2,
\end{equation}
where the operator $d^\dagger_{n\sigma}$ creates a dot electron 
in a single-particle state with orbital quantum number $n=1,\ldots,
M$ and spin projection $\sigma=\uparrow,\downarrow$.
The $2M\times 2M$ Hermitian matrix $h_{n\sigma,n'\sigma'}$ 
encapsulates the single-particle content, including SOC and 
magnetic field effects.  At this stage, we make no 
assumptions about the SOC, allowing
for rather general statements regarding the anomalous Josephson effect.  
Importantly, the $h$ matrix can always be 
diagonalized by a unitary transformation,
$U^\dagger h U = {\rm diag}(E_\nu)$, with the single-particle energies 
$E_\nu$ ($\nu=1,\ldots,2M$).
We then have associated fermionic operators, $c_\nu$, with
\begin{equation} \label{dvsc}
d_{n \sigma} = \sum_{\nu=1}^{2M} U_{n \sigma, \nu} \ c_\nu,
\end{equation}
which correspond to single-particle eigenstates of the
isolated dot. The $d_{n\sigma}$ operators instead will be taken to represent 
dot fermion modes tunnel-coupled to the leads.  
Both representations are, of course, equivalent, and the benefits of using
the $c_\nu$ should become clear below. After the unitary transformation,
\begin{equation}\label{hd2}
H_d = \sum_\nu E_\nu c^\dagger_\nu c^{}_\nu + 
E_c ( \hat N - n_g)^2.
\end{equation}
The capacitive Coulomb charging term is only sensitive to 
the total dot fermion number operator, 
\begin{equation}
\hat N=\sum_{n\sigma} d_{n\sigma}^\dagger d_{n\sigma}^{}=
\sum_\nu c^\dagger_\nu c^{}_\nu,
\end{equation}
where the charging energy, $E_c$, sets the energy cost 
for adding or removing electrons.
The real number $n_g$ is proportional to a backgate voltage and 
regulates the average number of electrons on the dot.
It is worth mentioning that the above charging term generically describes the 
dominant interaction contribution.\cite{nazarovbook} 
For later use, we also define the Coulomb 
energy differences $W_k$ (integer $k$), 
\begin{equation}\label{wk}
W_k = E_c(N_0 + k -n_g)^2 - E_c(N_0-n_g)^2 ,
\end{equation}
where the integer $N_0$ denotes the ground-state electron number on the dot. 

The left and right ($j=L,R$) superconducting
leads are described by standard bulk BCS Hamiltonians.
For simplicity, we assume that they have identical 
gap $\Delta$ and normal-state dispersion relation $\xi_{\bf k}$,
with chemical potential $\mu_S = 0$.\cite{footnote2} 
Moreover, we use a gauge where the order parameter phases appear in the 
tunneling Hamiltonian $H_t$ only, and $\Delta\ge 0$ is real-valued.  
It is then convenient to switch to particle-hole (Nambu) space and 
introduce the spinor $\Psi_{j{\bf k}}= ( c_{j,{\bf k},\uparrow}^{} ,
 c^\dagger_{j,-{\bf k},\downarrow})^T$, where
$c_{j,{\bf k},\sigma}^\dagger$ creates an electron in lead $j$ with 
momentum ${\bf k}$ and spin projection $\sigma$.
The lead Hamiltonian is then given by 
\begin{equation}\label{bcs}
H_l = \sum_{j=L,R}\sum_{\bf k} \Psi_{j {\bf k}}^\dagger 
\left ( \begin{array}{cc} \xi_{\bf k} & \Delta
\\ \Delta & -\xi_{\bf k}\end{array}\right) \Psi_{j {\bf k}}^{}.
\end{equation}
Finally, we come to $H_t$, where a complex-valued tunneling matrix element 
$t_{j,{\bf k},\sigma;n,\sigma'}$ gives the probability amplitude for 
transfer of an electron from  dot state $(n,\sigma')$ to 
lead state $(j,{\bf k},\sigma)$. 
 To simplify the analysis, we adopt the standard
wide-band approximation for the leads\cite{nazarovbook} and
neglect the ${\bf k}$-dependence of the tunneling matrix elements. 
Leaving aside spin-active interfaces,
tunneling is assumed to be spin-conserving and spin-independent, 
$t_{j,{\bf k}, \sigma; n \sigma'}= \delta_{\sigma \sigma'} t_{j,n}$,
and $H_t$ is determined by $2M$ complex-valued parameters $t_{j,n}$. 
Employing the Nambu spinor notation also for the dot fermions,
$D_{n} = (d_{n,\uparrow}^{}, d^\dagger_{n,\downarrow})^T$, we obtain
\begin{eqnarray}\label{ht}
H_t &=& \sum_{j=L,R} \sum_{\bf k}
\sum_{n=1}^M \Psi_{j{\bf k}}^\dagger T_{j,n} D_n + \textrm{H.c.},
\\ \nonumber 
T_{j,n} &=& \left( \begin{array}{cc} e^{i\phi_j/2} t_{j,n} & 0 \\ 
0 & - e^{-i\phi_j/2} t_{j,n}^\ast \end{array} \right),
\end{eqnarray}
where $\phi_j$ is the superconducting phase in lead $j$.

\subsection{Current-phase relation}

In this paper, we study the equilibrium Josephson CPR in the
zero-temperature limit, $T\to 0$.  A formally
exact expression for the CPR can be obtained 
from the partition function, $Z={\rm Tr}e^{-\beta H}$, with $\beta=1/T$. 
We start by employing Wick's theorem to trace out the non-interacting
lead fermions.  In the interaction picture, let $H_0=H-H_t$
govern the imaginary-time ($\tau$) evolution.
For arbitrary operator ${\cal O}$, we use the notation\cite{abrikosov}
\begin{equation} \label{rule}
{\cal O}(\tau)= e^{H_0\tau} {\cal O} e^{-H_0\tau} , \quad
\bar{\cal O}(\tau)= e^{H_0\tau} {\cal O}^\dagger e^{-H_0\tau} .
\end{equation}
The partition function then reads
\begin{eqnarray}\nonumber
Z &=& {\rm Tr}_d {\rm Tr}_l \left( e^{-\beta H_0} {\cal T}
e^{-\int_0^\beta d\tau  H_t(\tau)}\right) \\ \label{zeff}
&=& Z_{l} {\rm Tr}_d \left ( e^{-\beta H_d} {\cal T}e^{-S_t} \right),
\end{eqnarray}
where ${\cal T}$ denotes time ordering. The traces ${\rm Tr}_{d,l}$ are
over dot and lead Hilbert spaces, respectively, with $Z_l=
{\rm Tr}_l e^{-\beta H_l}$.  In Eq.~\eqref{zeff}, we 
have averaged over the leads, and using
$\langle H_t(\tau)\rangle_l^{} = Z_l^{-1} {\rm Tr}_l [
 e^{-\beta H_l} H_t (\tau) ] = 0,$
Wick's theorem implies that $S_t$ in Eq.~\eqref{zeff} 
is completely determined by the Gaussian correlator
\begin{equation}
S_t = -\frac12 \int_0^\beta d\tau d\tau'
\left \langle {\cal T} H_t(\tau) H_t(\tau')\right\rangle_l^{}.
\end{equation}
Inserting $H_t$ [Eq.~\eqref{ht}], we obtain
\begin{equation} \label{st}
S_t = \frac12 \int_0^\beta d\tau d\tau' \sum_{nn'} \bar D_n(\tau) 
\Lambda_{nn'}(\tau-\tau') D_{n'}^{}(\tau'),
\end{equation}
where $\Lambda_{nn'}(\tau-\tau') = 2 \sum_j 
 T^\dagger_{j,n} G_l(\tau-\tau')  T_{j,n'}$
is expressed in terms of the lead Green's function,
\begin{eqnarray} 
G_l(\tau-\tau') &=& -\sum_{\bf k} \left\langle {\cal T} \Psi_{j{\bf k}}(\tau)
\bar\Psi_{j{\bf k}}(\tau')\right\rangle_l \\ \nonumber
&=& -\pi\nu_0 T\sum_m \frac{e^{-i\omega_m(\tau-\tau')}}
{\sqrt{\omega_m^2+\Delta^2}}
\left( \begin{array}{cc}i\omega_m &  \Delta \\ \Delta & i\omega_m \end{array}
\right),
\end{eqnarray}
which is identical for both leads.  Here we have employed the 
wide-band approximation, with normal-state lead density of states 
$\nu_0=\sum_{\bf k} \delta(\xi_{\bf k})$,
and fermion Matsubara frequencies $\omega_m=\pi T (2m+1)$ 
(integer $m$).  The kernel $\Lambda$ in Eq.~(\ref{st}),
describing the effects  of the traced-out leads on the dot fermions, thus reads
\begin{equation} \label{Lambda}
\Lambda_{nn'}(\tau)  =  \sum_{j=L,R} \Gamma_{n n'}^{(j)}
\left( \begin{array}{cc} \partial_\tau & \Delta e^{- i \phi_j } \\
\Delta e^{i\phi_j} & \partial_\tau \end{array}  \right) f(\tau),
\end{equation}
where the tunnel contacts are described by Hermitian $M\times M$ 
hybridization matrices,
\begin{equation}\label{hybmat}
\Gamma_{nn'}^{(j)}= 2 \pi \nu_0 t_{j, n}^\ast t^{}_{j, n'},
\end{equation}
and we use the auxiliary function
\begin{equation}\label{fdef}
f(\tau) = T \sum_{m} \frac{e^{-i\omega_m\tau}}{\sqrt{\omega_m^2 + \Delta^2}} .
\end{equation}
Notice that $\Lambda$ factorizes in orbital and Nambu subspaces.

The Josephson current flowing through contact $j$ to the dot follows
from the ground-state average\cite{jj}
\begin{equation}\label{sup}
I_j = \frac{2e}{\hbar} \partial_{\phi_j} F,
\end{equation}
where $F=-T\ln Z$ is the free energy.
Current conservation dictates $I_{L,R} = \pm I(\varphi)$, 
where $\varphi = \phi_L - \phi_R$ is
the gauge-invariant phase difference.   Using Eqs.~(\ref{zeff}) and 
(\ref{sup}), the $T=0$ CPR, $I(\varphi)$, will be
computed  in Sec.~\ref{sec3} for the cotunneling regime and in the
atomic limit.

\subsection{Two orbital levels} \label{sec:model}

For concrete results, we will consider a generic
 model with $M=2$ dot orbital levels, which provides a minimal 
setting for studying SOC effects, the anomalous supercurrent, 
and Majorana fermions.
The $4\times 4$ matrix $h$ describing the single-particle spectrum
of the dot Hamiltonian $H_d$ [Eq.~\eqref{hd}] is taken in the generic form
\begin{equation} \label{h}
h = (\mu \tau_0 + \epsilon \tau_z) \sigma_0 + 
B \tau_0 \sigma_z + \alpha \tau_y \left[
\cos(\chi) \sigma_z + \sin (\chi) \sigma_y \right]  ,
\end{equation}
where $\tau_{x,y,z}$ ($\sigma_{x,y,z}$) are Pauli matrices 
in orbital (spin) space;
the respective unity matrices are $\tau_0$ ($\sigma_0$).
The physics is here determined by the interplay of a Rashba-type SOC, whose strength is parameterized by the energy scale $\alpha$, and
the magnetic Zeeman field, with energy scale $B$. In Eq.~\eqref{h},
$0\le \chi\le \pi$ denotes the angle 
between the effective spin-orbit field and the Zeeman field.
The bare [$\alpha=B=0$] dot levels are $\mu \pm \epsilon$. 
For the specific 2D dot model studied in Ref.~\onlinecite{prbSO}, 
it is straightforward to explicitly determine
the model parameters entering Eq.~\eqref{h}.  

Next we express the $2\times 2$ (in orbital space) hybridization matrices
[Eq.~\eqref{hybmat}] in the form
\begin{equation} \label{Gammaj}
\Gamma^{(j=L,R)} = \gamma_j \left( \begin{array}{cc} 
e^{\lambda_j} & e^{i \delta_j}
\\ e^{- i \delta_j} & e^{-\lambda_j} \end{array} \right) ,
\end{equation}
where $\gamma_j\ge 0$ gives the overall hybridization strength
of the respective contact, $\lambda_j$ parametrizes the
orbital asymmetry (for $\lambda_j=0$, both orbitals couple 
symmetrically to the $j$th lead),  
and $\delta_j$ is an inter-orbital phase shift.
Since $\delta_{L,R}$ is independent of spin, these phase shifts have 
nothing to do with SOC. For instance, they could be caused by 
orbital magnetic fields; for the dot model proposed in Ref.~\onlinecite{prbSO}, 
this follows by virtue of a gauge transformation 
transferring the orbital field dependence to the
tunneling Hamiltonian.  The phases $\delta_{L,R}$ may
also be influenced by the dot geometry, in particular by 
contact asymmetries.  It is worth stressing that 
for $\alpha\ne 0$ and $\Delta\ne 0$,
one cannot gauge away the resulting phases $\delta_{L,R}$.  
For further convenience, we define the relative inter-orbital phase shift 
\begin{equation}\label{delta}
\delta= \delta_L-\delta_R.
\end{equation}
In the absence of SOC, i.e., for $\alpha=0$, the dot Hamiltonian is
diagonal in orbital space, and then only the phase difference
$\delta$ cannot be gauged away.

We note that our assumption of ${\bf k}$-independent tunneling matrix
elements implies that the phase shifts $\delta_j$
are also momentum-independent. If this assumption is violated, 
the $\delta_j$ are best treated as statistical variables.
The resulting average may suppress $I_a$ while leaving 
critical currents basically unaffected.  
Since such generalizations are straightforward to implement, 
we here proceed by assuming ${\bf k}$-independent 
phase shifts $\delta_{L,R}$.

\section{Josephson current} \label{sec3}

In this work, we compute the Josephson current for 
the above model using two complementary
vantage points, namely by perturbation theory in the cotunneling
regime and by employing an effective Hamiltonian valid in the deep 
subgap regime. 

\subsection{Cotunneling regime}\label{sec3a}

The cotunneling regime is realized when all eigenvalues
of the hybridization matrices $\Gamma^{(L,R)}$ are small against $\Delta$.  
In that case, perturbation theory in these Hermitian
matrices is well-defined and allows for progress.\cite{footcot}
Since $S_t\propto \Gamma^{(L,R)}$, see Eqs.~\eqref{st} and \eqref{Lambda},
the free energy $F$ can be directly expanded in powers of $S_t$.
Starting from Eq.~(\ref{zeff}) and using 
$\partial_{\phi_j}\langle S_t\rangle =0$, the lowest-order 
contribution to the Josephson current \eqref{sup} is of order 
$\Gamma^L\Gamma^R$ and reads
$I_j = -2T \left\langle S_t \partial_{\phi_j}S_t\right \rangle,$
where $\langle \cdots\rangle$ denotes the ground-state
expectation value for the closed dot Hamiltonian $H_d$.
Inserting Eq.~(\ref{st}), we find $I_{L,R}=\pm I(\varphi)$,
in accordance with current conservation, where 
\begin{equation}\label{cotun}
I(\varphi) = I_0 \sin\varphi+ I_a \cos\varphi
\end{equation}
with the currents 
\begin{eqnarray} \label{i0a}
\left( \begin{array}{c} I_0 \\ i I_a \end{array}\right) &=&
\sum_{nmn'm'} 
\left(\begin{array}{c} 
\Gamma^{(L)}_{nm} \Gamma^{(R)}_{n'm'} + (L\leftrightarrow R) \\
\Gamma^{(L)}_{nm} \Gamma^{(R)}_{n'm'} - (L\leftrightarrow R) 
\end{array}\right) 
\\ &\times& \nonumber
 \frac{\Delta^2}{2\beta} 
\int_0^\beta d \tau_1 d \tau_2 d \tau'_{1} d \tau'_{2} 
\ f (\tau_1 - \tau_2 ) f (\tau_{1}' - \tau_{2}' ) 
\\ \nonumber &\times&
\left\langle {\cal T} d_{n \downarrow}(\tau_1) d_{m\uparrow}(\tau_2)
\bar d_{n' \uparrow} (\tau_{1}') \bar d_{m' \downarrow} (\tau_{2}')
\right\rangle .
\end{eqnarray}
The critical current is $I_c=\sqrt{I_0^2+I_a^2}$, where 
we find $I_{c1}=I_{c2}=I_c$ in the cotunneling regime.
It is now crucial to use the unitary transformation $U$ in  
Eq.~\eqref{dvsc} to switch from the $d_{n\sigma}$ to the $c_\nu$ fermions.
The latter represent the eigenstates of the isolated interacting dot.  
Using $f(\tau) = f(-\tau)$, we observe that only the antisymmetric part 
of the transformed hybridization matrices enters the expressions for $I_{0,a}$.
In terms of the antisymmetric $2M\times 2M$ matrices
\begin{equation} \label{tildeGj}
\tilde\Gamma_{\nu \mu}^{(j=L,R)} =  \sum_{nm} \Gamma_{nm}^{(j)} \left(
 U_{n \downarrow, \nu} U_{m \uparrow, \mu} - 
 U_{n \downarrow, \mu} U_{m \uparrow, \nu}\right),
\end{equation}
we find from Eq.~(\ref{i0a}) for the anomalous Josephson current
\begin{equation}\label{Ica}
I_a = \frac{e\Delta^2}{\hbar} \sum_{\nu>\mu}  J_{\nu\mu} \ Q_{\nu \mu},
\end{equation}
with the symmetric $2M\times 2M$ matrices
\begin{equation}\label{Jdef}
J_{\nu\mu} =  {\rm Im} \left( \tilde \Gamma_{\nu \mu}^{(L)}
[\tilde \Gamma^{(R)}]^\ast_{\nu \mu} \right),
\end{equation}
\begin{eqnarray} \nonumber
Q_{\nu \mu} &=& - T \int_0^\beta d \tau_1 d \tau_2 d \tau_{1}' d \tau_{2}'
 \ f \left(\tau_1 - \tau_2 \right) f \left(\tau_{1}' - \tau_{2}' \right) \\
& &\times  \left\langle {\cal T} c^{}_\nu (\tau_1) c_\mu^{} (\tau_2)
 \bar c_\nu (\tau_{1}') \bar c_\mu(\tau_{2}') \right\rangle.
\label{Qnumu}
\end{eqnarray}
The current $I_0$ follows in similar form,
\begin{equation}\label{Icc}
I_0 =  \frac{e\Delta^2}{\hbar} \sum_{\nu>\mu}
{\rm Re}\left (\tilde \Gamma_{\nu \mu}^{(L)} \ 
[\tilde \Gamma^{(R)}_{\nu\mu}]^\ast \right) \ Q_{\nu \mu}.
\end{equation}

We can now use Eq.~(\ref{Ica}) to infer general conditions for
the anomalous Josephson effect to exist within the cotunneling regime.
As \textit{necessary condition} for $I_a\ne 0$, we observe
that  $J_{\nu\mu}\ne 0$ must be satisfied 
for at least one index pair $\nu>\mu$.
Note that $J_{\nu\mu}$ depends only on single-particle
quantities, such as tunneling matrix elements, SOC, and Zeeman fields.
The role of interactions is encoded in the $Q$ matrix and
can be crucial in breaking the balance between time-reversed processes, 
which may then induce the anomalous Josephson effect. 
Note that this condition is very general and holds for arbitrary
matrices $h$ determining the single-particle spectrum.

It is interesting to see what happens
for a single-level dot, $M=1$, where $\Gamma^{(L)}$ and $\Gamma^{(R)}$
are just real numbers. The antisymmetric $\tilde \Gamma^{(L,R)}$ matrices 
in Eq.~\eqref{tildeGj} are then fully determined by 
$\tilde \Gamma^{(j)}_{21}=  \Gamma^{(j)} ( U_{\downarrow,2}
U_{\uparrow,1} - U_{\downarrow,1}U_{\uparrow,2} )$,
which immediately yields $J=0$ in Eq.~(\ref{Jdef}).  Hence no anomalous
Josephson current is possible in a single-orbital dot, even when
interactions are included.  A minimal model for this effect has
to start from $M=2$ orbital dot levels, see Sec.~\ref{sec4}, where 
we study the conditions for the anomalous Josephson effect in a concrete
and experimentally relevant setting.

General conditions (beyond the cotunneling regime)
for the anomalous Josephson effect can 
also be deduced directly from symmetry considerations.
We exemplify this here by analyzing the supercurrent through 
an inversion-symmetric two-dimensional dot with in-plane (purely Zeeman) 
magnetic field $B$ and SOC strength $\alpha$.
A spatial inversion operation, $(x,y)\to (-x,-y)$, is implemented 
by (i) exchanging the lead indices, $L\leftrightarrow R$, (ii)
inverting the phase difference, $\varphi \to - \varphi$,  (iii)
changing the sign of the SOC, $\alpha \to - \alpha$, and (iv) 
also changing the sign of the (in-plane) Zeeman field, $B\to -B$.  
Since $I(\varphi) \to - I(-\varphi)$ under spatial inversion, 
Eq.~\eqref{cotun} implies that the 
anomalous supercurrent must satisfy the symmetry relation 
\begin{equation}\label{symmetry}
I_a\left (\Gamma^{(L)}, \Gamma^{(R)}, B, \alpha\right) =
 -I_a \left(\Gamma^{(R)}, \Gamma^{(L)},- B, -\alpha\right) .
\end{equation}
Similarly, we deduce an additional 
condition from the supercurrent behavior under a time reversal operation, 
\begin{equation}
I_a\left (\Gamma^{(L)}, \Gamma^{(R)}, B, \alpha\right) = -
I_a \left(\Gamma^{(L)}, \Gamma^{(R)},- B, \alpha\right) ,
\end{equation}
which implies that $I_a$ is always odd in $B$.

Let us next address the $Q$ matrix in Eq.~\eqref{Qnumu},
which only depends on properties of the closed dot. In the cotunneling regime,
interactions can affect the CPR only through this matrix.  
In general, $4!=24$ terms involving
all possible permutations of time-ordered fermion operators will be generated
from Eq.~\eqref{Qnumu}.
However, if the closed dot has a non-degenerate interacting 
ground state $|G\rangle$, Eq.~\eqref{Qnumu}
allows for simplifications in the $\beta\to \infty$ limit
of interest here.  Excluding 'accidental' degeneracies, 
this step assumes that a TRS-breaking magnetic field is present.
Effectively, only three permutations in Eq.~\eqref{Qnumu} 
are relevant and $Q_{\nu\mu}$ can be expressed in terms of the
three real-valued functions
\begin{eqnarray}\nonumber
{\cal Q}_i(\epsilon_a,\epsilon_b,\epsilon_c) &=&
\frac{1}{\beta} \int_0^\beta d \tau_a \int_0^{\tau_a} d \tau_b \int_0^{\tau_b}
d \tau_c \int_0^{\tau_c} d \tau_d \\ &\times& \label{Q2}
e^{-\epsilon_a (\tau_a - \tau_b) -\epsilon_b (\tau_b - \tau_c)
-\epsilon_c (\tau_c - \tau_d)} \\ &\times& \nonumber \left\{ \begin{array}{ll}
f(\tau_a - \tau_b) f(\tau_c - \tau_d), & i = 1, \\ \\
f(\tau_a - \tau_d) f(\tau_b - \tau_c), & i = 2, \\  \\
f(\tau_a - \tau_c) f(\tau_b - \tau_d),  & i = 3, \end{array} \right. 
\end{eqnarray}
where $\epsilon_{a,b,c}\ge 0$ are possible excitation energies.
Switching to the frequency domain and using Eq.~\eqref{fdef}, 
we obtain\cite{foot4}
\begin{eqnarray} \label{Q123}
&& {\cal Q}_i = \int \frac{d \omega_1 d \omega_2}{(2 \pi)^2} \,
\frac{1}{\sqrt{(\omega_1^2+\Delta^2)(\omega_2^2+\Delta^2)}}
\\ \nonumber &&\times \left\{ \begin{array}{ll} 
(1 - \delta_{\epsilon_b, 0})/
[(i \omega_1 + \epsilon_a )( i \omega_2 + \epsilon_c ) \epsilon_b] ,& i=1,\\ \\
1/[( i \omega_1 + \epsilon_a)( i \omega_1 + \epsilon_c )
(i\omega_1 + i\omega_2 + \epsilon_b)] ,& i=2,\\ \\
1/[( i \omega_1 + \epsilon_a )( i \omega_2 + \epsilon_c )
( i \omega_1 + i\omega_2 + \epsilon_b) ] ,& i=3.\end{array} \right. 
\end{eqnarray}
Notice that the ${\cal Q}_i$ are invariant under the exchange 
$\epsilon_a\leftrightarrow \epsilon_c.$ 
Consider now the ground state $|G\rangle$ of the closed
dot Hamiltonian $H_d$ in Eq.~(\ref{hd2}), with
$N_0$ electrons on the dot, $\hat N|G\rangle = 
N_0|G\rangle$. Assuming that $|G\rangle$ is non-degenerate,
the filling factor $n_\nu$ for each single-particle state 
$\nu=1,\ldots,2M$ is known.  Arranging the $E_\nu$ as ordered sequence, 
$E_1 \leq E_2 \leq \cdots\leq E_{2M}$, the result is
\begin{equation}\label{nnu}
n_\nu = \langle G | c_\nu^\dagger c_\nu^{} | G \rangle =
\left\{ \begin{array}{ll} 1, & \nu\le N_0, \\ 0,&\nu>N_0. \end{array}\right.
\end{equation}
For given index pair $\nu> \mu$, three possibilities arise, namely
$(n_\nu,  n_\mu ) = (0,0)$, $(1,1)$, and $(0,1)$.  It is then
straightforward to determine the excitation energies 
$\epsilon_{a,b,c}$ by comparing Eqs.~\eqref{Qnumu}
and \eqref{Q2} in those three cases.  
To state the final result for $Q$, it is useful to
introduce the positive energies
\begin{eqnarray}\label{enumu}
\tilde E_\nu &=& (1-2n_\nu) E_\nu + W_{1-2n_\nu},\\ \nonumber
\tilde E_{\nu\mu} &=& (1-2n_\nu)E_\nu+ (1-2n_\mu)E_\mu+W_{2-2n_\nu-2n_\mu} ,
\end{eqnarray}
with the Coulomb energy differences $W_k$ in Eq.~\eqref{wk}.
(Note that for $E_c=0$, we have $1-2n_\nu={\rm sgn}(E_\nu)$ and hence
$\tilde E_\nu=|E_\nu|$.)  We then obtain the symmetric $Q$ matrix,
\begin{eqnarray}  \nonumber
Q_{\nu \mu} &=& (1-2n_\nu)(1-2 n_\mu) \Bigl[ 
2{\cal Q}_{i_1}(\tilde E_\nu , \tilde E_{\nu\mu}, \tilde E_\mu) \\
&+& \label{calQ} 
 {\cal Q}_{i_2}( \tilde E_\nu, \tilde E_{\nu\mu}, \tilde E_\nu  ) 
\\ \nonumber
&+& {\cal Q}_{i_2}( \tilde E_\mu, \tilde E_{\nu\mu}, \tilde E_\mu  ) 
+ 2 {\cal Q}_{i_3}( \tilde E_\nu ,0, \tilde E_\mu ) \Bigr],
\end{eqnarray}
where the indices are $i_1=i_2=1$ and $i_3=3$ for $n_\nu=n_\mu$.
For $n_\nu\ne n_\mu$, we instead have $i_1=i_3=2$ and $i_2=3$.  

We proceed by discussing the limit of strong Coulomb blockade.
For $E_c \to \infty$, the cotunneling supercurrent is generally strongly 
suppressed.  Technically, this suppression can be seen 
from Eq.~\eqref{Q123}: all excitation energies scale as
$\epsilon_{a,b,c}\propto E_c\to \infty$, which implies $Q_{\nu \mu} \to 0$ 
and thus $I_{0,a}\to 0$.  This argument only breaks down for
half-integer values of $n_g$, where the strong charging term in $H_d$ allows
for two degenerate charge states with particle numbers $N_0=
N_{0,\pm} \equiv n_g\pm 1/2$.  Let us therefore now focus on half-integer 
values of $n_g$, where the single-particle spectrum, $\{E_\nu\}$, 
ultimately determines the ground state and, in particular,
which particle number $N_0$ is realized (either $N_{0,+}$ or $N_{0,-}$).  
Using that for $N_0=N_{0,\pm}$, we have the Coulomb energy difference 
$W_{\mp 1}=0$, Eq.~\eqref{calQ} simplifies to 
\begin{eqnarray} \label{Qinf}
&& Q_{\nu \mu}^{(N_{0,+})} = 2 n_\nu n_\mu 
 {\cal Q}_3 ( -E_\nu, 0, -E_\mu ) \\ \nonumber && - \left[( 1 - n_\nu) n_\mu
{\cal Q}_3( -E_\mu, E_\nu - E_\mu, -E_\mu ) +(\nu\leftrightarrow\mu)\right],
 \\ \nonumber &&
Q_{\nu \mu}^{(N_{0,-})} = 2 ( 1 - n_\nu ) ( 1 - n_\mu )
{\cal Q}_3 ( E_\nu, 0, E_\mu )\\ \nonumber && -
\left[ ( 1 - n_\nu ) n_\mu 
{\cal Q}_3 (E_\nu, E_\nu - E_\mu, E_\nu ) +(\nu \leftrightarrow \mu)\right].
\end{eqnarray}
It is instructive to examine Eq.~\eqref{Qinf} for a spin-degenerate 
single-level ($M=1$) dot without SOC and without magnetic field.  
Both single-particle states ($\nu=\uparrow$, $\downarrow$)
then have identical energy, say $E_\nu= x\Delta$ with some
dimensionless parameter $x$, 
and Eq.~\eqref{Qinf} yields\cite{footnote1}
\begin{equation}\label{pi}
 Q_{\uparrow\downarrow} = {\cal Q}_3 (|x|\Delta,0,|x|\Delta)
 \times \left\{ \begin{array}{ll} 2,  & 
N_0 =0,2,  \\ - 1, & N_0 =1,  \end{array} \right.
\end{equation}
where Eq.~\eqref{Q123} gives ($x>0$)
\begin{equation}\label{q3}
{\cal Q}_3(x\Delta,0,x\Delta)= \frac{1}{\pi^2\Delta^3}
\frac{(\pi/2)^2 (1-x)- {\rm Arccos}^2 x}{x(1-x^2)}.
\end{equation} 
Noting that $I_a=0$ for $M=1$, the critical current $I_c$ directly
follows from Eq.~\eqref{Icc}, where Eq.~(\ref{pi}) predicts
$\pi$-junction [$0$-junction] 
behavior, with $I(\varphi)=-I_c\sin\varphi$ 
[$I(\varphi)=I_c\sin\varphi$], for $N_0=1$ [$N_0=0,2$].
We have thereby reproduced well-known results.\cite{jj,alfredo}
In general, in the strong Coulomb blockade limit $E_c\to \infty$, 
we find $\pi$-junction behavior for odd $N_0$ and half-integer $n_g$.

\subsection{Superconducting atomic limit}\label{sec3b}

We now turn to the atomic limit, where $\Delta$ represents the largest 
relevant energy scale and we can effectively put $\Delta \to \infty$. 
This allows us to go beyond the perturbative cotunneling regime and to 
compute the free energy $F$ without further approximations.
Using $f(\tau) \to \Delta^{-1} \delta(\tau)$ 
in Eq.~\eqref{Lambda}, the partition function reads $Z={\rm Tr}_d
e^{-\beta H_{\rm eff}}$.  The 'effective dot Hamiltonian' is 
\begin{equation} \label{Heff}
H_{\rm eff}  =  H_d + \frac12 \sum_{j=L,R}\sum_{n m} 
\left( \Gamma^{(j)}_{n m} e^{i \phi_j} d_{n \downarrow} d_{m \uparrow} + 
{\rm H.c.} \right),
\end{equation}
with $H_d$  in Eq.~\eqref{hd} and a proximity-induced 
$s$-wave pairing term due to the traced-out superconducting leads.\cite{alfredo}
The CPR then follows from Eq.~\eqref{sup}.
Notice that the Hilbert space of the dot can now be decomposed into
two independent sectors with even and odd fermion parity, respectively.

Equation \eqref{Heff} can be used to demonstrate that 
already in the cotunneling regime the limits $E_c \to \infty$ and 
$\Delta \to \infty$ do not commute.  For $\Delta\to\infty$,
one needs to retain only those contributions in Eq.~\eqref{Qnumu}
where two fermions forming a Cooper pair are tunneling as a whole, 
with the correlator of the form
$\langle {\cal T} c_\nu (\tau + 0^{+}) c_\mu (\tau)
\bar c_\nu (\tau' + 0^{+}) \bar c_\mu (\tau')\rangle$.
Using $\tilde E_{\nu\mu}\ge 0$ in Eq.~\eqref{enumu}, 
some algebra gives
\begin{equation}\label{qatlim}
Q_{\nu \mu} = \frac{\delta_{n_\nu , n_\mu}}{2 \Delta^2} \
\frac{1 - \delta_{\tilde E_{\nu\mu}, 0}}{\tilde E_{\nu\mu}}.
 \end{equation}
Since now $Q_{\nu\mu}\ge 0$ for arbitrary $N_0$,
$\pi$-junction behavior is never possible in the atomic limit,
in contrast to what we found for $\Delta<E_c\to \infty$ above.
This statement always applies within the atomic limit,
i.e., also beyond the cotunneling regime.
Moreover, in the atomic limit, $E_c<\Delta\to\infty,$
current flows only in the vicinity of the $2e$-charge 
degenerate points, where $W_{\pm 2} = 0$ in Eq.~\eqref{wk},
corresponding to integer values of $n_g$.
This again differs from the strong-blockade result in Eq.~(\ref{Qinf}),
where current flows only for half-integer $n_g$.
We thus conclude that the limits $E_c \to \infty$ and
$\Delta \to \infty$ do not commute.

\section{Anomalous Josephson current} \label{sec4}

In this section, we address the CPR and, in particular,
the anomalous supercurrent, $I_a=I(\varphi=0)$,
for the two-level dot in Sec.~\ref{sec:model}.

\subsection{Cotunneling regime}

In the cotunneling regime, the currents $I_{0}$ and $I_a$ 
determining the Josephson CPR [Eq.~\eqref{cotun}] follow from 
Eqs.~(\ref{Icc}) and (\ref{Ica}), respectively. 
The anomalous supercurrent is expressed in terms of the $4\times 4$
matrices $J$ and  $Q$, see Eqs.~\eqref{Jdef} and \eqref{calQ}, respectively,
where a necessary condition for the anomalous Josephson effect is given
by $J_{\nu\mu}\ne 0$ for at least one index pair $\nu>\mu$.  
In order to evaluate the $J$ matrix, we need the unitary matrix $U$ 
diagonalizing $h$.  While $U$ is easily found, the lengthy
result is not illuminating; we use it only to obtain the data shown
in the figures below. Instead, we will provide analytical results in 
several complementary limits,
where the algebra is simpler and allows for an intuitive picture.

Let us first observe that when the spin-orbit field is perpendicular 
to the Zeeman field ($\chi = \pi/2$), $h$ is a 
symmetric matrix. Hence the diagonalizing matrix $U$ can 
always be chosen to have only real-valued entries, 
and for $\delta_{L,R}=0$, we obtain $J=0$ from Eq.~\eqref{Jdef}. 
We conclude that for $\chi=\pi/2$, the anomalous Josephson effect is 
only possible when at least one of the phase shifts $\delta_{L,R}$ 
is non-zero.  This conclusion is in accordance with previous work.\cite{prlSO}

\subsubsection{Collinear spin-orbit and Zeeman fields}

\begin{figure}[t]
\centering
\includegraphics[width=7cm]{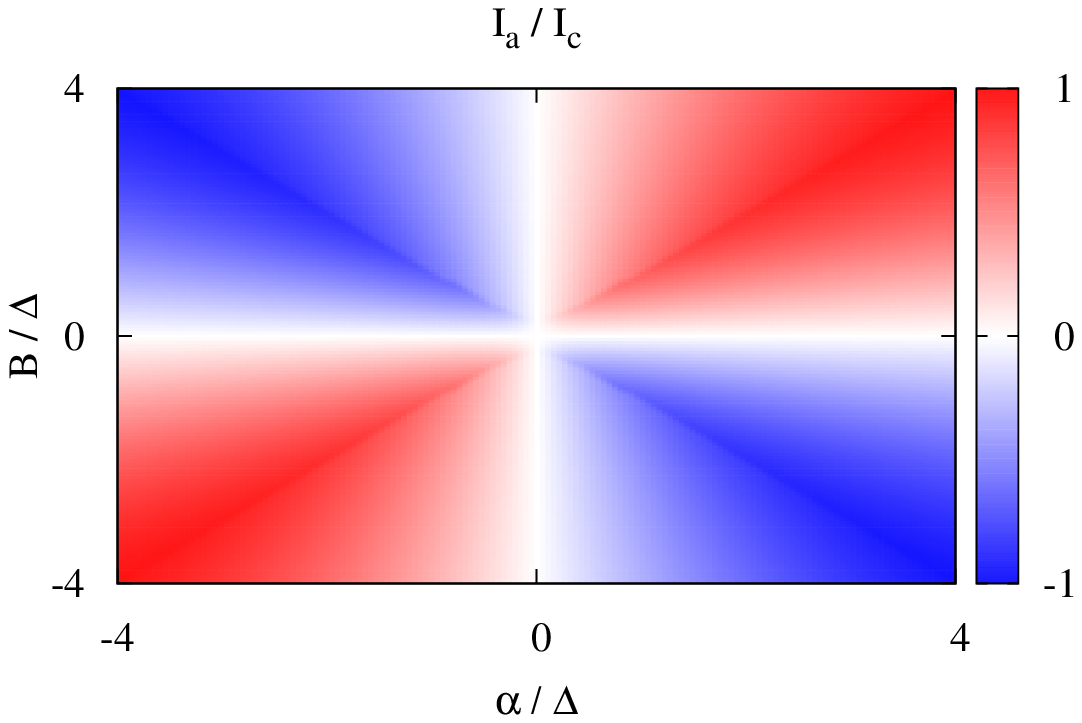} 
\includegraphics[width=8cm]{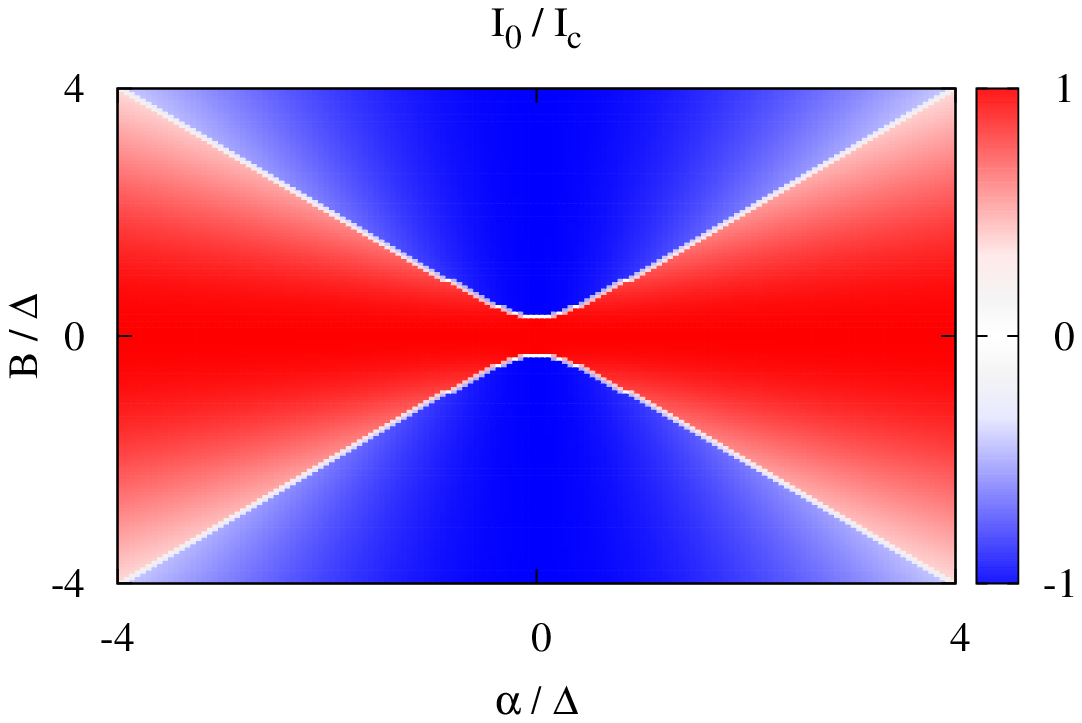}
\caption{\label{fig1} 
Anomalous supercurrent ($I_{a}$, top panel) and 'normal' supercurrent
 ($I_0$, bottom) determining the cotunneling CPR \eqref{cotun}
in the $B$-$\alpha$ plane.   The results are for the two-level dot 
with $\epsilon=0.3\Delta$, $E_c=2\Delta$, $n_g=2$, and 
$\chi=\delta_{L,R}=\mu=\lambda_L=0$. 
For the right contact, only the orbital level $n=1$ is assumed to couple
to the superconductor, i.e., $\lambda_R\to \infty$ with 
$\gamma_R e^{\lambda_R}\to \gamma_R$.
Note that $I_{a,0}$ are normalized to the respective critical
current $I_c= \sqrt{I_0^2+I_a^2}$.  }
\end{figure}

\begin{figure}[t]
\centering
\includegraphics[width=8cm]{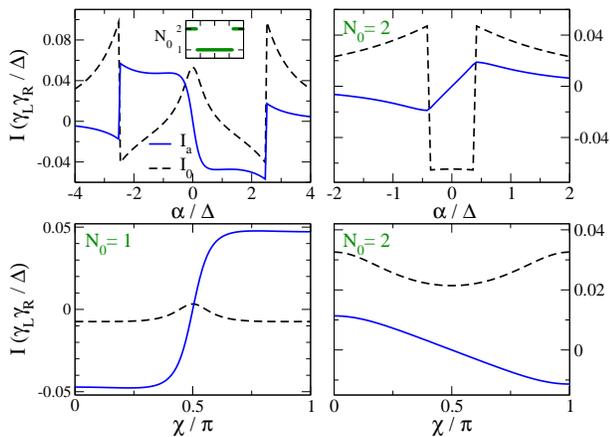}
\caption{\label{fig2} 
Parameter dependence of $I_{a,0}$ 
(main panels) and of the particle number $N_0$ (inset) for
 $B=0.5\Delta$, with other parameters as in Fig.~\ref{fig1}.
Blue solid curves show $I_a$, and black dashed curves $I_0$, 
both in units of $e\gamma_L\gamma_R/\hbar\Delta$.
Top row:  SOC $\alpha$ is varied for fixed field angle $\chi=0$, 
with $n_g=1$ (left) and $n_g=2$ (right).  
Bottom row:  $\chi$ is varied for fixed $\alpha=1.2\Delta$, 
with $n_g=1$ (left) and $n_g=2$ (right).  }
\end{figure}

\begin{figure}[t]
\centering
\includegraphics[width=8cm]{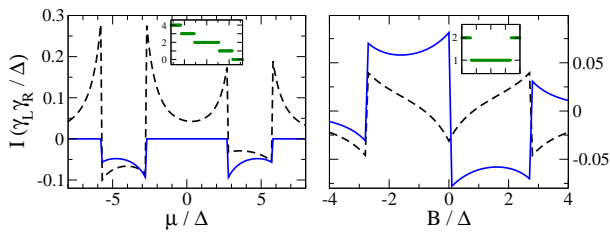}
\caption{\label{fig3} 
Same as Fig.~\ref{fig2} but showing 
$I_{a,0}$ vs $\mu$ for $B=0.001\Delta$ (left), and $I_{a,0}$ vs $B$ for 
$\mu=3\Delta$ (right). Other parameters are
as in Fig.~\ref{fig1} except for $E_c=1.5\Delta$.  }
\end{figure}

\begin{figure}[t]
\centering
\includegraphics[width=8cm]{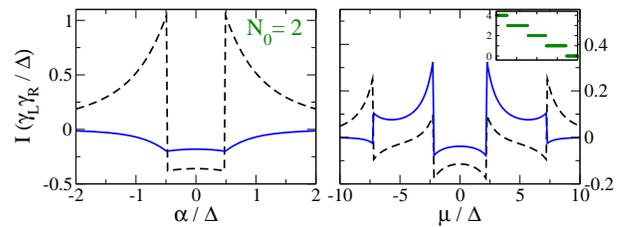}
\caption{\label{fig4}
Same as Fig.~\ref{fig2} but showing
$I_{a,0}$ vs $\alpha$ for $\mu=E_c=0$ (left), and  $I_{a,0}$ vs $\mu$
for $\alpha=0$ and $E_c=2\Delta$ (right). We use the parameters 
$\epsilon=0.5\Delta$, $B=0.7\Delta$, $n_g=2$, 
$\chi=  \lambda_{L,R}= \delta_R=0$, and $\delta_{L}=\pi/2$.  }
\end{figure}

{}From Ref.~\onlinecite{prlSO},  we then expect that the anomalous 
supercurrent is maximal for $\chi = 0$, 
where spin-orbit and Zeeman fields point along the same direction.
We thus consider $h$ in Eq.~(\ref{h}) for $\chi=0$, where
the diagonalization matrix is 
\begin{equation} \label{unita}
U = e^{i \tau_x \sigma_z \theta / 2}, \quad \sin\theta=\frac{ \alpha}{E_d},
\quad E_d=\sqrt{\epsilon^2+\alpha^2},
\end{equation}
and the spectrum $(E_1,\ldots,E_4)$ is given by
$\mu+(E_d+B,E_d-B,-E_d+B,-E_d-B)$.
Using Eq.~\eqref{tildeGj}, the antisymmetric hybridization 
matrices $\tilde \Gamma^{(L,R)}$ have the nonvanishing entries
\begin{eqnarray*} 
\tilde \Gamma^{(j)}_{23} &=& -[\tilde \Gamma^{(j)}]^\ast_{14}  = 
\gamma_j \left( \cos\delta_j  + i
 \frac{ \alpha\sinh\lambda_j + \epsilon \sin \delta_j}{E_d}  \right),\\
\tilde \Gamma^{(j)}_{21} &=& \tilde \Gamma^{(j)}_{43}\Big|_{\theta 
\to \theta + \pi} = 
\gamma_j\left(\cosh\lambda_j + \frac{ \epsilon\sinh\lambda_j-
 \alpha \sin\delta_j}{E_d} \right).
\end{eqnarray*}
The symmetric $J$ matrix in Eq.~\eqref{Jdef} thus has 
the non-zero elements
\begin{equation} \label{calI23chi}
J_{32} =\frac{\gamma_L\gamma_R}{E_d} \left [ \epsilon\sin \delta 
+ \alpha  ( \cos\delta_R\sinh\lambda_L-\cos\delta_L
\sinh\lambda_R ) \right]
\end{equation}
and $J_{41}=-J_{32}$.  
Remarkably, this result does not depend on the Zeeman field $B$.  In the end, the anomalous supercurrent is
\begin{equation} \label{Ia23}
I_a = \Delta^2 J_{32} \left( Q_{32} - Q_{41} \right) .
\end{equation}
Several observations can be drawn from the above equations.

First, note that $J_{32}=0$ for $\Gamma^{(L)} = \Gamma^{(R)}$ (where
$\delta_L=\delta_R$ and $\lambda_L=\lambda_R$). Therefore,
asymmetric tunnel contacts with matrices $\Gamma^{(L)}\ne 
\Gamma^{(R)}$ are necessary for $I_a\ne 0$, see Ref.~\onlinecite{prlSO}.
The resulting typical 'phase diagram' for $I_{a,0}$ in the $B$-$\alpha$ plane 
is depicted in Fig.~\ref{fig1}.
The standard Josephson effect, where one has either $0$- or $\pi$-junction
behavior with $|I_a/I_0|\ll 1$, is recovered
when either $\alpha$ or $B$ are small.  In contrast, the anomalous supercurrent 
is most pronounced when $|\alpha|\approx |B|$. 
The lower panel (for $I_0$) indicates that within the Zeeman-dominated
regime $|B|>\sqrt{\alpha^2+\epsilon^2}$, we have $I_0<0$, implying that
 $\pi$-junction behavior can be realized.   Furthermore, we observe that 
for the chosen parameter set, $I_a$ is odd in the product $\alpha B$.

The $\alpha$-dependence for fixed $B=0.5\Delta$ is shown in
the upper panel of Fig.~\ref{fig2}.  The steps in $I_{0,a}$ vs
$\alpha$ (and in all
figures below) can be traced back to level degeneracies, 
where higher-order perturbative
terms become important and will smear out the steps.
For the chosen parameters and $n_g=2$, we have $N_0=2$ for all shown SOCs, 
but for $n_g=1$ (upper left panel), $N_0=1$ for certain $\alpha$.
The anomalous supercurrent is generally enhanced for odd $N_0$
compared to the even-$N_0$ case.

The lower-row panels in Fig.~\ref{fig2} show the $\chi$-dependence
of $I_{a,0}$ for SOC $\alpha=1.2\Delta$, confirming that the
anomalous supercurrent is maximized for $\chi=0$~mod~$\pi$ but vanishes
for $\chi=\pi/2$.   In addition, by comparing to the respective
$E_c=0$ plots (not shown), we observe that $I_a$ is not 
drastically affected by interactions while $I_0$ becomes suppressed.
This suggests that interactions tend to enhance the relative
importance of the anomalous supercurrent.

Next we observe that in general $Q_{32} \neq Q_{41}$. As long as 
$J_{32}\ne 0$, an anomalous supercurrent may then flow. 
This could happen for arbitrary (including zero) SOC
$\alpha$. However, we always need a finite Zeeman field. Indeed,
for $B=0$, we find that $Q_{32}=Q_{41}$ due to 
level degeneracies ($E_1=E_2$ and $E_3=E_4$), and hence $I_a=0$ for 
$B=0$, cf.~also Fig.~\ref{fig1}. Nonetheless, anomalous
supercurrents can survive even for arbitrarily weak $B$,
in particular when interactions are present.
We will address this issue in more detail below
for the resonant case ($\epsilon=0$),
but Fig.~\ref{fig3} already illustrates the phenomenon
for $\epsilon=0.3\Delta$. The left panel in Fig.~\ref{fig3}
shows that even for $B=0.001\Delta$, in the presence of interactions
and with odd $N_0$, the anomalous supercurrent is finite and sizeable.
Similarly, the right panel shows that for $B\to 0$, we obtain
an unusual $I_a(B)$ dependence instead of the 
standard linear $B$-dependence discussed in Ref.~\onlinecite{prlSO}.
We expect that higher-order perturbative corrections smear out
the cusps near $B=0$, see also Sec.~\ref{sec4b},
 and eventually lead to $I_a\propto {\rm sgn}(B)$.

Let us now analyze the case without SOC:
Putting $\alpha=0$ in Eq.~\eqref{calI23chi}, we observe 
that $I_a\ne 0$ is possible for relative inter-orbital phase 
shift $\delta \neq 0$, cf.~Eq.~\eqref{delta}. 
The possibility of an anomalous Josephson effect 
induced by the magnetic field alone (without SOC) 
in a noninteracting multi-level dot was overlooked in 
Ref.~\onlinecite{prlSO}, where only the case $\delta_{L,R} = 0$ has
been studied.   This effect is shown in Fig.~\ref{fig4} for 
phase shifts $\delta_R=0$ and $\delta_L=\pi/2$ (otherwise the
tunnel contacts are here assumed identical, $\lambda_L=\lambda_R$).
The left panel, where $N_0=2$ for the chosen parameters, 
illustrates the counter-intuitive increase in $|I_a|$ as
the SOC is decreased.  In fact, here we find the largest possible anomalous
supercurrent for $\alpha=0$.
Note that, as a consequence of the inter-orbital phase shift $\delta=\pi/2$,
the anomalous supercurrent is now an even function of the SOC parameter
 $\alpha$.  The right panel presents the $\mu$-dependence of $I_a$,
where we see again that the anomalous supercurrent
is enhanced whenever $N_0$ is odd.   

Finally, let us note that for $\delta=0$, the condition $J_{32} \neq 0$, 
with Eq.~\eqref{calI23chi} for $J_{32}$, is equivalent to
$\alpha \neq 0$ and  nonvanishing commutator 
$\left[ \Gamma^{(L)}, \Gamma^{(R)} \right] \neq 0$, which 
corresponds to the chirality condition in Sec.~\ref{sec1}.
 These two necessary conditions for anomalous  supercurrents
were specified in Ref.~\onlinecite{prlSO}. 

\subsubsection{Resonant level} \label{sec:resonant}

Another interesting and nontrivial situation emerges when the
two bare levels are resonantly aligned.  Then
$\epsilon=0$ (with arbitrary $\chi$) in Eq.~\eqref{h},
and the unitary matrix $U$ diagonalizing $h$ is 
\begin{equation}\label{unitary1}
U= e^{i \tau_x \pi / 4} e^{i  \hat \theta \sigma_x/ 2},
\end{equation}
where $\hat\theta={\rm diag}(\theta_+,\theta_-)$ is a 
diagonal matrix in orbital space.  The angles $\theta_\pm$ follow from
\begin{equation}\label{phipm}
e^{i\theta_\pm} = \frac{B \pm e^{i\chi}\alpha}{E_\pm},
\quad E_\pm = \sqrt{\alpha^2 +B^2 \pm 2 \alpha B \cos \chi},
\end{equation}
and $(E_1,\ldots,E_4)=\mu+(E_+,-E_+,E_-,-E_-)$.  Some algebra shows that 
the symmetric $J$ matrix has the nonvanishing elements\cite{foot5}
\begin{equation} \label{calI12eps}
J_{21} = \gamma_L\gamma_R\left( \cos\delta_L
\sinh\lambda_R-\cos\delta_R\sinh\lambda_L \right)
\end{equation}
and $J_{43}=-J_{21}$. 
 For the anomalous Josephson current, we thus find 
\begin{equation} \label{Ia12}
I_a = \Delta^2 J_{21} \left( Q_{21} - Q_{43} \right) .
\end{equation}
Quite remarkably, $J_{21}$ in Eq.~\eqref{calI12eps} 
neither depends on the Zeeman field $B$ nor on the SOC
$\alpha$.  In principle, we may then expect $I_a\ne 0$ 
even for very small $\alpha$ and/or  $B$.  In addition, $J_{21}$ 
does not depend on $\chi$ either, and it is not obvious why
$I_a=0$ for $\chi=\pi/2$ as discussed above.
However, we also need to examine the contribution of the $Q$ matrix.
In fact, when $\alpha B \cos\chi=0$, the level degeneracy $E_+=E_-$ implies
from Eq.~\eqref{calQ} that $Q_{21} = Q_{43}$, which in turn 
gives $I_a=0$ for $\epsilon=0$ and arbitrary $E_c$.

\begin{figure}[t]
\centering
\includegraphics[width=8cm]{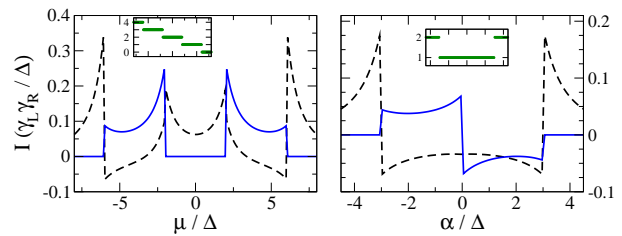}
\caption{\label{fig5} 
Same as Fig.~\ref{fig2} but for the resonant orbital ($\epsilon=0$) case
with tiny Zeeman field, $B=0.001\Delta$.  The left panel shows
$I_{a,0}$ vs $\mu$ for $\alpha=0.001\Delta$, while the right 
panel displays $I_{a,0}$
vs $\alpha$ for $\mu=5\Delta$. 
The remaining parameters are as in Fig.~\ref{fig1}.  }
\end{figure}

\begin{figure}[t]
\centering
\includegraphics[width=8cm]{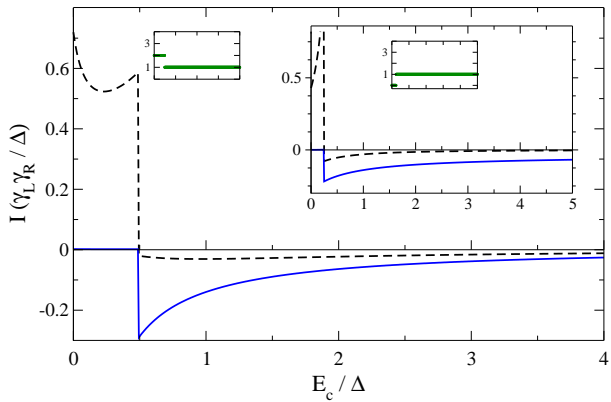}
\caption{\label{fig6} 
Same as Fig.~\ref{fig2} but showing $I_{a,0}$ vs $E_c$
for $n_g=2$ (main panel) and $n_g=3/2$ (large right inset), with
$\alpha=0.5\Delta$, $B=0.01\Delta$, and $\epsilon=0.01\Delta$.  }
\end{figure}

Nonetheless, we again encounter the possibility that  $I_a \neq 0$ 
even for very small Zeeman field $B$ and temperatures $T<|B|$,
suggesting the incipient spontaneous breakdown of TRS (note that TRS is 
restored for $B=0$). Remarkably, this onset behavior can be triggered by
Coulomb interactions even for very small SOC $\alpha$.
Before going through the detailed argument, we first illustrate
this behavior for $B=0.001\Delta$ in Fig.~\ref{fig5}.
The left panel indeed reveals a finite and sizeable anomalous
supercurrent for $\alpha=B=0.001\Delta$ if interactions are present,
$E_c\ne 0$, and $N_0$ is odd.  The right panel suggests
that $I_a\propto {\rm sgn}(\alpha B)$ for arbitrarily
small (but finite) $\alpha$.  For $\alpha=0.5\Delta$, the interaction
effects in this interesting parameter regime are 
displayed in Fig.~\ref{fig6}. While $I_a=0$ for small $E_c$,
we find $I_a\ne 0$ for $E_c\agt |\alpha|$, with
$|I_a|$ weakly decreasing in the limit of strong Coulomb blockade.
For the resonant case of half-integer $n_g$, $I_a$ saturates
at a finite value for $E_c\to \infty$, cf.~inset of Fig.~\ref{fig6}.

Next we aim at understanding the above $I_a\propto {\rm sgn}(\alpha B)$
onset behavior.  To simplify the algebra as much as possible, we put $\chi=0$ and
consider the limiting case of very small but finite $(B,\alpha)$,
where interactions play a crucial role.
(For $|\alpha|\gg |B|$, the arguments below 
show that the onset behavior $I_a\propto {\rm sgn}(B)$  
is possible even when $E_c=0$.) Equation \eqref{phipm} then gives 
$e^{i\theta_\pm}=\pm {\rm sgn}(\alpha)$ for $|\alpha|>|B|$, and thus the
complex-valued unitary matrix in Eq.~\eqref{unitary1} 
has different limits for positive and negative SOC,
$\lim_{\alpha\to 0^+} U \ne \lim_{\alpha\to -0^+} U$.
This corresponds to different residual 'magnetizations' of the
$\tau\otimes\sigma$ isospin near the $SU(4)$ 
symmetric point in parameter space defined by $B = \alpha = 0$.
(Note that in the absence of hysteresis, $I_a=0$ directly 
at the symmetric point, since then $U={\rm diag}(1)$ is 
real-valued and thus implies $J=0$.) 
Recall next that the columns of $U$ are eigenvectors of $h$, 
forming four linearly independent isospin projections.  
The corresponding single-particle energy levels
are $ \mu+ \{ |\alpha|+ \eta, - |\alpha| - \eta, |\alpha| - \eta, 
- |\alpha|+\eta \}$ with $\eta= {\rm sgn}(\alpha) B$.
When $\mu$ is chosen such that $N_0 =1$, assuming $B>0$,
one spin-$\downarrow$ electron will occupy the
single-particle level $E_2$ ($E_4$) for $\alpha > 0$ ($\alpha < 0$).
For $N_0 = 1$, we observe that
${\rm sgn} (Q_{21}) = - {\rm sgn} (Q_{43}) = -
 {\rm sgn}(\alpha)$, see Eq.~\eqref{calQ} with ${\cal Q}_i > 0$, and therefore
 Eq.~\eqref{Ia12} suggests that we may have a finite anomalous supercurrent. 
However, for very small $(B,\alpha)$ and $E_c=0$, the 
energy separation between states with different $N_0$ is also tiny.
This eventually results in the complete cancellation of all
time-reversed contributions, and $I_a=0$ in the noninteracting 
case for very small $B$ and $\alpha$.  For finite charging energy, however,
the energy gap to states with different $N_0$ grows with $E_c$,
which  renders the $N_0=1$ ground state more robust.
Taking the small-$(B,\alpha)$ limit for finite $E_c$ should then
leave ground-state properties such as $N_0$ or the spin polarization
unaffected, and $I_a \propto {\rm sgn}(\alpha B)$ remains finite.
However, the above arguments also show that $I_a$
will be suppressed by thermal fluctuations once the temperature
scale exceeds the Zeeman field scale. Therefore the 
$I_a\propto{\rm sgn}(\alpha B)$ onset behavior just found for the ground state 
can 'only' be interpreted as incipient breakdown of TRS, i.e.,
TRS is restored by thermal fluctuations for $T>|B|$. 

Analytical results for the ground-state anomalous supercurrent 
are possible in the strong Coulomb blockade limit.
For instance, at the charge degeneracy point $n_g=3/2$ with $N_0=1$, 
Eq.~\eqref{Ia12} yields for small $(B,\alpha)$ the result
\begin{equation}\label{anomalous}
I_a = - 3 \ {\rm sgn}(\alpha B)  \Delta^2 J_{21} {\cal Q}_3(\mu, 0, \mu ),
\end{equation}
where $J_{21}$ and ${\cal Q}_3(\mu,0,\mu)$ are given in 
Eqs.~\eqref{calI12eps} and \eqref{q3}, respectively.
This confirms explicitly the $I_a \propto {\rm sgn}(\alpha B)$ onset behavior
discussed above.

\subsection{Superconducting atomic limit}\label{sec4b}

\begin{figure}[t]
\centering
\includegraphics[width=8cm]{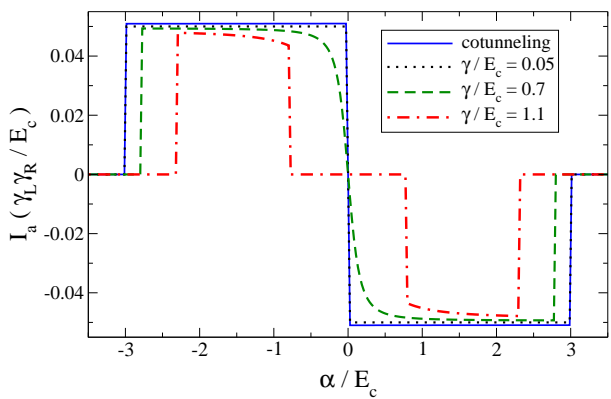}
\caption{\label{fig7} 
Anomalous supercurrent $I_a$ for the two-orbital 
dot vs SOC $\alpha$ in the atomic 
limit ($\Delta\to \infty$) for several $\gamma=\gamma_L=\gamma_R$.
The shown results follow from Eq.~\eqref{AT2}
and the effective dot Hamiltonian (\ref{Heff}).
The other parameters are as in the right panel 
of Fig.~\ref{fig5}: $\epsilon=0$, $B/E_c=0.0005$, $\mu/E_c=2.5$, $n_g=2$, 
$\chi=\delta_{L,R}=\lambda_L=0$, and $\lambda_R\to \infty$.  
The solid blue curve gives the respective cotunneling result 
[Eq.~\eqref{Ia12} with $\Delta\to \infty$] for $\gamma/E_c=0.05$.  }
\end{figure}

Next we briefly turn to a discussion of the anomalous Josephson effect in the
 superconducting atomic limit, see Sec.~\ref{sec3b}, 
where the $\Delta\to \infty$ 
effective dot Hamiltonian, $H_{\rm eff}$ in Eq.~\eqref{Heff},
allows us to go beyond the perturbative cotunneling regime.
Evaluating the anomalous Josephson current at the, say, 
left contact, we obtain
\begin{equation} \label{AT2}
I_a = -\frac{2e}{\hbar} {\rm Im} \sum_{\nu < \mu} 
\tilde\Gamma_{\nu\mu}^{(L)} \left \langle c_\nu c_\mu \right\rangle ,
\end{equation}
where the brackets indicate a ground-state average using 
$H_{\rm eff}(\varphi=0)$. 
We consider the two-orbital dot in Sec.~\ref{sec:model}, where
the $4\times 4$ hybridization matrices $\tilde\Gamma^{(L,R)}$ follow from 
Eq.~\eqref{Gammaj} after transformation to the $c_\nu$ fermion representation.
As detailed in Sec.~\ref{sec3b}, the $\varphi$-dependent
ground-state energies should be computed separately for
the (decoupled) odd and even fermion parity subspaces. 
We then expect $I_a\ne 0$ only when the ground state (for $\varphi=0$)
has odd parity.  

The dependence of $I_a$ on the SOC $\alpha$ is illustrated in Fig.~\ref{fig7},
where we use parameters as in the right panel of Fig.~\ref{fig5}.
This allows us to study how the $I_a\propto {\rm sgn}(\alpha B)$ onset behavior
(the signature of incipient TRS breaking) emerges
from the cusp features encountered in perturbation theory.
First, we note from Fig.~\ref{fig7} that the cotunneling result (taking
$\Delta\to \infty$ in the above expressions) matches the 
predictions of Eq.~\eqref{AT2} for $\gamma_{L,R}\to 0$. 
This matching has also been confirmed analytically by perturbative
expansion of the general $\Delta\to \infty$ cotunneling result
[see Eqs.~\eqref{Ica} and \eqref{qatlim}] to lowest nontrivial order
in the hybridization matrices.  We conclude that the 
limits $\gamma_{L,R}\to 0$ and $\Delta\to\infty$ commute.  
Second, cusp-like features as seen in the right panel of Fig.~\ref{fig5}
emerging under a perturbative theory will be smeared out by higher-order
corrections, and indeed imply $I_a\propto{\rm sgn}(\alpha B)$
onset behaviors associated with time-reversal symmetry breaking.  
Third, for large hybridizations $\gamma_{L,R}$, it is also 
possible that the fermion parity of the resulting $\varphi=0$ ground state 
is changed.  This is apparent in Fig.~\ref{fig7}, where we find $I_a=0$ for
small $|\alpha|$ and $\gamma/E_c=1.1$ as 
a consequence of such a transition.
The anomalous supercurrent can here be tuned to  
zero either by raising $\gamma$ or by lowering $E_c$.

\section{Majorana fermions}\label{sec5}

We proceed by noting that  all ingredients needed for the realization of 
Majorana fermions\cite{carlo,alicea} are in principle present 
in our model, namely
proximity-induced superconductivity, SOC, and a TRS-breaking
magnetic field.  As discussed below, the Majorana regime can be reached
in the superconducting atomic limit of the two-level dot in 
Sec.~\ref{sec:model}, where the two orbitals here correspond to two
spatially separated single-level dots (i.e., a double dot).
The resulting MBSs are topologically unprotected, i.e.,
their realization requires the fine-tuning of gate voltages, 
Zeeman field $B$, and/or phase difference $\varphi$.
Gate voltages here affect the orbital asymmetry
 $\epsilon$ through confinement potentials, 
the average energy $\mu$, and/or the SOC $\alpha$.
For a spatially separated MBS pair -- such that both
MBSs correspond to different orbital states, allowing to distinguish 
them -- we find characteristic signatures in the $2\pi$-periodic
CPR. This is in marked contrast to the 'fractional' $4\pi$-periodic CPR for
 topologically protected Majoranas,\cite{alicea}  which
has not been observed so far due to difficulties in ensuring fermion
parity conservation in practice.

We use the atomic-limit effective Hamiltonian
 $H_{\rm eff}$  in Eq.~\eqref{Heff} for the double dot. 
Using the basis $\{ | 1, \uparrow \rangle , | 2, \downarrow \rangle, 
| 1, \downarrow \rangle, | 2, \uparrow \rangle \}$,
the single-particle matrix $h$ [Eq.~\eqref{h}] has the representation
\begin{equation}\label{hnew}
h = \left( \begin{array}{cccc}
\mu+\epsilon + B & - \alpha \sin\chi & 0 & i \alpha \cos \chi \\
- \alpha \sin \chi & \mu-(\epsilon + B) & i \alpha \cos \chi & 0 \\
0 & - i \alpha \cos \chi & \mu+\epsilon - B & \alpha \sin \chi \\
- i \alpha \cos \chi & 0 & \alpha \sin \chi & \mu-(\epsilon - B)
\end{array} \right) .
\end{equation}
Without losing generality, $\alpha>0$ and $B>0$ from now on. We
approach a suitable parameter regime
 by comparing to the Kitaev chain\cite{qi,carlo,alicea,karsten}
describing 1D (effectively spinless) $p$-wave topological superconductors.
The Kitaev chain is known to support MBSs, and 
based on this analogy we choose $\chi = \pi / 2$, i.e.,
Zeeman and spin-orbit fields are perpendicular.\cite{alicea}  
$h$ is then block-diagonal
with decoupled upper and lower two-state subspaces.
 The connection to the Kitaev chain becomes clear
when $\epsilon$ is positive and chosen in the parameter regime
\begin{equation}\label{mbsreg}
\Delta\gg \epsilon + B \gg {\rm max}\left (\alpha,  \left|\epsilon - B\right|, 
\gamma_{L,R},\mu,E_c \right).
\end{equation}
The upper-block state $(2, \downarrow)$ will then always be occupied, while
 $(1,\uparrow)$ is always empty.  The upper left block in Eq.~\eqref{hnew} 
can thus be projected away, and the resulting truncated Hamiltonian,
$H_{\rm eff}'$, acts only within the lower right block described by 
the (effectively spinless) fermion operators
$d_1 \equiv d_{1 \downarrow}$ and $d_2\equiv  d_{2 \uparrow}$,
\begin{eqnarray}\nonumber
H_{\rm eff}' &=& (\mu+\epsilon-B)d_1^\dagger d_1^{} +
[\mu-(\epsilon-B)]d_2^\dagger d_2^{}\\  &+&
E_c \left ( d_1^\dagger d_1^{}+ d_2^{\dagger} d_2^{}-n_g\right)^2 
\label{mf:H}\\ \nonumber &+& \left( \alpha d^\dagger_1 d^{}_2 +
 \tilde\Delta(\varphi) e^{i\vartheta(\varphi)} 
d^\dagger_2 d^\dagger_1 + {\rm H.c.}\right),
\end{eqnarray}
where the occupied $(2,\downarrow)$ state leads to a shift 
$n_g\to n_g+1$. 
With the hybridization matrix (\ref{Gammaj}), the double dot model
in Eq.~\eqref{Heff} yields the complex-valued effective pairing amplitude
$\tilde \Delta e^{i\vartheta} = 
\frac12 \sum_{j} \gamma_j e^{-i (\phi_j + \delta_j)}$.
It is now convenient to introduce $\gamma\equiv (\gamma_L+\gamma_R)/2$,
and to gauge away the overall phase $\sum_j (\phi_j + \delta_j)/2$. 
We then obtain 
\begin{eqnarray}\label{gammad}
\tilde\Delta (\varphi)&=& \gamma 
\sqrt{1-T_0 \sin^2[(\varphi+\delta)/2]},\\ \nonumber
&&\quad T_0 =  \frac{ 4\gamma_L\gamma_R}{
(\gamma_L+\gamma_R)^2}, \\ \nonumber
\vartheta(\varphi) &=& \tan^{-1}\left(\frac{\gamma_R-\gamma_L}
{\gamma_R+\gamma_L}   \tan[(\varphi+\delta)/2] \right), 
\end{eqnarray} 
with the phase shift $\delta$ in Eq.~\eqref{delta}.
Note that $0\le T_0\le 1$ corresponds to the transmission 
probability of a single-channel quantum point contact, while
$\tilde\Delta(\varphi)$ gives the Andreev level energy in 
the atomic limit.\cite{nazarovbook}

We proceed by first discussing the noninteracting case, $E_c=0$, 
where two spatially resolved MBSs may appear when the (necessary) conditions
\begin{equation}\label{mbs1cond}
 B = \epsilon,\quad \mu = 0
\end{equation} 
are met. $H_{\rm eff}'$ can then be diagonalized 
in terms of fermionic Bogoliubov-de Gennes (BdG) quasiparticle operators,
\begin{equation} \label{mf:eta}
\eta_\pm = \frac12 \left[ d_1 + d_2 \pm  e^{i\vartheta}
\left( d^\dagger_1 - d^\dagger_2 \right) \right],  
\end{equation}
where Eq.~\eqref{mf:H} yields the BdG Hamiltonian
\begin{equation} \label{mf:H2}
H_{\rm eff}' = \sum_{\pm} E_\pm(\varphi) \left( \eta^\dagger_\pm 
\eta^{}_\pm - \frac12 \right) , \quad
E_{\pm} = \alpha \pm \tilde\Delta(\varphi).
\end{equation}
The four possible single-particle eigenstates are constructed by applying 
$\eta^\dagger_\pm$ or $\eta_\pm^{}$ to the vacuum state, with the
respective energies $E_\pm/2$ and $- E_\pm/2$.
The CPR then follows from Eq.~\eqref{mf:H2},
\begin{equation}
I(\varphi) = 2 \frac{ \partial\tilde\Delta}{\partial\varphi}
\ [ \Theta (- E_+)- \Theta(-E_-)],
\end{equation}
where $\Theta$ is the Heaviside function. 
Notice that $I=0$ for $\tilde\Delta(\varphi)<\alpha$, 
since both energies $E_{\pm}=\alpha\pm \tilde\Delta$ have the same sign.  
We therefore find
\begin{eqnarray} \label{i0phi}
I(\varphi) &=& \Theta( \tilde\Delta - \alpha ) 
I_0(\varphi), \\ \nonumber  I_0(\varphi) &=& \frac{e\gamma}{2\hbar} 
\frac{T_0 \sin( \varphi+\delta) }{\sqrt{1-T_0\sin^2[(\varphi+\delta)/2]}},
\end{eqnarray}
where $I_0(\varphi)$ 
coincides with the CPR of a single-channel quantum point contact with 
transparency $T_0$,\cite{alfredo} shifted by the
inter-orbital phase difference $\delta$.
The CPR (\ref{i0phi}) is $2 \pi$-periodic in $\varphi$ and 
vanishes (or reappears) at the boundaries between 
ground states with opposite fermion parity. These boundaries 
are precisely the formation points of MBSs, as we show next.

Noting that both $\alpha$ and $\tilde\Delta$ are non-negative,
 the zero-energy condition for MBS formation is satisfied for 
$E_-(\varphi)=0$, i.e., for
\begin{equation}\label{mbs2cond}
\tilde\Delta(\varphi)= \alpha.
\end{equation}
This corresponds to a pair of zero-energy MBSs,
generated by the anticommuting Majorana fermion operators
$\xi_1= -i ( \eta^{}_- - \eta^\dagger_-)$ and 
$\xi_2= \eta^{}_- + \eta^\dagger_-$; 
note that $\xi_n=\xi_n^\dagger$ and $\xi_n^2=1$. 
 In order to avoid recombination to a conventional 
fermion, we need both MBSs to be spatially separated.
This is achieved for 
\begin{equation}\label{mbs3cond}
\vartheta(\varphi)=0 \ {\rm mod}\ \pi,
\end{equation}
where $\xi_{1}$ and $\xi_2$ have well-defined and 
different orbital quantum numbers, and thus correspond to different
single-level dots in this double dot.
Taking for instance $\vartheta=0$, Eq.~\eqref{mf:eta} yields
$\xi_1 =-i( d_1{}-d^{\dagger}_{1} )$ and $\xi_2 = d^{}_{2} + d^\dagger_{2}$,
which indeed implies that the MBS associated with $\xi_{n=1 \ (2)}$  
has the orbital wavefunction  $n=1 \ (2)$. 
We conclude that Eq.~\eqref{mbs3cond} ensures that both
 MBSs are spatially separated.  Using Eq.~\eqref{gammad}, 
there are two possibilities to satisfy this condition:
(1) We may choose equal hybridization strengths,
$\gamma_L=\gamma_R=\gamma$. Then
$T_0=1$, which implies $\tilde\Delta=\gamma|\cos[(\varphi+
\delta)/2]|=\alpha$, with two solutions (for $\varphi$) when
$\gamma>\alpha$. For these two phase values, MBSs will be present.
(2) Alternatively, for $\gamma_L\ne \gamma_R$, another
possibility emerges by adjusting $\varphi=-\delta$ (mod $2\pi$),  
where Eq.~\eqref{mbs2cond} allows for a MBS pair when $\gamma=\alpha$.
Realizing either of those conditions amounts to reaching the 'sweet spot' 
for a Kitaev chain with two fermion sites, see also
Refs.~\onlinecite{flensberg,wright,fulga}. 
The MBS solutions are here quadratically protected against
small deviations in the effective dot levels, see Eq.~(\ref{mbs1cond}).
While there is no such protection against deviations from the condition
(\ref{mbs2cond}), this lack of protection also offers the advantage of 
MBS tunability by variation of the superconducting phase difference.
Noting that already for a three-site chain, robust 
protection of unpaired MBSs can be achieved,\cite{wright}
we expect that a reasonable compromise between well protected
MBSs and good tunability is possible using our double dot
proposal.  

Let us now see how the above scenario will be affected by weak
electron-electron interactions.  
We here continue to use the 'global charging
energy' in Eq.~(\ref{hd}), since for a double dot 
in the large-$B$ limit of interest here, see Eq.~(\ref{mbsreg}), 
both dots are effectively occupied by one fermion at most.
In that case, the global charging energy is equivalent to a capacitive
inter-dot interaction.  For finite $E_c$, the system can be tuned
to the MBS regime by replacing the condition $\mu=0$ in Eq.~\eqref{mbs1cond}
by $\mu = - 2E_c ( 1-n_g )$, i.e., by putting 
$\mu$  at the charge degeneracy point. ($B=\epsilon$ is still
required.)
In terms of the $\eta_{\pm}$ operators in Eq.~\eqref{mf:eta}, 
the Hamiltonian \eqref{mf:H} then reads
\begin{equation} \label{mf:H3}
H = \sum_{ \pm} E_\pm(\varphi) \left( \eta^\dagger_\pm \eta^{}_\pm 
-\frac12 \right) + E_c \left( \eta^\dagger_+ \eta^{}_+- 
\eta^\dagger_- \eta^{}_- \right)^2 ,
\end{equation}
with $E_\pm(\varphi)$ in Eq.~\eqref{mf:H2}.
The MBS regime is realized when there are two 
ground states with opposite fermion parity.
By examining the many-particle spectrum of Eq.~\eqref{mf:H3}, 
\begin{eqnarray}\label{manyspec}
E_{0,0} &=& - \alpha, \quad  E_{1,0} =\tilde\Delta+ E_c,\\ \nonumber
E_{0,1} &=& -\tilde\Delta + E_c ,\quad  E_{1,1} = \alpha ,
\end{eqnarray}
where $E_{n_+,n_-} $ denotes the energy of a state with 
$n_\pm = \langle \eta^\dagger_\pm \eta_\pm^{} \rangle$,
the condition (\ref{mbs2cond}) for the appearance of MBSs
is replaced by
\begin{equation}\label{boc}
\alpha = \tilde\Delta(\varphi) - E_c >0 .
\end{equation}
In the MBS regime, one has a double-degenerate ground state,
corresponding to negative energy eigenvalues $E_{0,1} = E_{0,0}$.
Inclusion of the charging energy thus only shifts the conditions
for Majorana formation, and below we focus on the case $E_c=0$.
Our proposal is therefore rather different from the double-dot scenario in 
Ref.~\onlinecite{flensberg}, where MBSs are induced
only in the limit of strong intra-dot Coulomb interactions while
the magnetic Zeeman field can be arbitrarily small.  

\begin{figure}[t]
\centering
\vspace*{1cm}
\includegraphics[width=8cm]{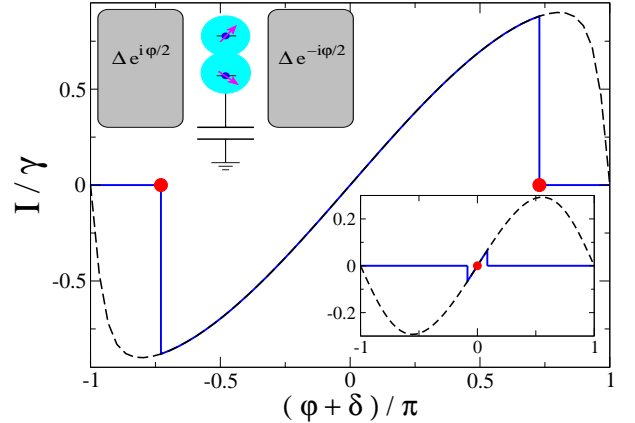}
\caption{\label{fig8} CPR through a double dot in the atomic limit
[see Eq.~\eqref{mbsreg}] with $B=\epsilon$, $\mu=E_c=0$, and $\chi=\pi/2$.
Main panel: CPR (blue solid curve) for $\alpha = 0.4 \gamma$, 
where $\gamma=(\gamma_L+\gamma_R)/2$ with slightly asymmetric
$\gamma_{R,L}$ such that $T_0 = 0.99$.
Red points on the CPR indicate that for the respective value of 
$\varphi$, a MBS pair is formed (see main text).
The dashed black curve shows the CPR for $\alpha=0$, where no MBSs occur.
The top left inset shows the schematic setup. The bottom right
inset gives the CPR for $\alpha=0.99\gamma$ and 
significant hybridization asymmetry, $T_0=0.5$, as blue solid curve. 
The red point indicates MBS pair formation, and the dashed curve
is for $\alpha=0$ (without MBSs).
}
\end{figure}

The Josephson current [Eq.~\eqref{i0phi}]   
turns out to be nonzero (zero) for odd (even) $N_0$, where the  
 CPR in general consists of two different regions:
For $\tilde\Delta(\varphi)>\alpha$, we find $I=I_0(\varphi)$ as for 
a single-channel quantum point contact (but with a phase shift
when $\delta\ne 0$), while $I=0$
for $\tilde\Delta<\alpha$.  At the boundary between both
regions, the parity $(-)^{N_0}$ changes 
from odd to even (or vice versa). It is precisely at these points that
two degenerate 'half-fermion' BdG quasi-particle states appear.
Under the described conditions, these can form a pair of spatially 
separated MBSs.
Observation of $I=0$ within a part of the CPR can then serve
as indirect signature for the MBSs, as illustrated
in Fig.~\ref{fig8}.  While jumps in the CPR can also
have a different origin, the peculiar feature linked to the appearance
of MBS pairs is the complete vanishing of the supercurrent in
a finite phase interval.  For the asymmetric case shown in the 
lower inset of Fig.~\ref{fig8}, the other two points on the CPR where 
the current vanishes correspond to spatially overlapping MBSs.

Ideally, one should thus consider a symmetric setup with $T_0=1$
in order to satisfy Eq.~(\ref{mbs3cond}).  The MBSs can then 
be detected through parity changes causing abrupt current jumps in the CPR.
In contrast, for asymmetric cases with $T_0<1$, Eq.~(\ref{mbs3cond}) 
is satisfied only at $\varphi=-\delta$,
where MBSs cannot be detected via transport measurements.
With decreasing transparency $T_0$, corresponding to increasing 
overlap between both MBSs [note that $\vartheta$
in Eq.~(\ref{gammad}) is a function of $T_0$],
the critical current decreases and the flat region $(I=0)$ in the CPR
gets shorter. In fact, Eq.~(\ref{i0phi}) predicts that for transparencies
$T_0<T_c$, with a critical transparency value determined by
$\gamma\sqrt{1-T_c^2}= \alpha$, there will be no flat CPR regions,
and hence no abrupt current jumps, at all.
Finally, it is worthwhile pointing out that in contrast to the
fractional Josephson effect for topologically protected Majoranas,\cite{alicea}
the MBSs discussed here do not mediate a Josephson current themselves.

\section{Conclusions}\label{sec6}

In this work, we have analyzed two particularly interesting
aspects of  Josephson transport in hybrid superconductor-dot systems
-- a pair of conventional BCS superconductors 
connected through a multi-level quantum dot -- where SOC, Coulomb
charging and magnetic field effects are taken into account.
First, we have studied the conditions for
the anomalous Josephson effect, i.e., supercurrent flow for vanishing
phase difference.  It is remarkable that Coulomb interactions can
qualitatively affect this phenomenon to allow for 
ground-state anomalous supercurrents
even when time-reversal breaking perturbations are very small compared
to all other relevant scales.
As described in Sec.~\ref{sec4}, 
we find spontaneously broken time-reversal symmetry with
anomalous supercurrent flowing for arbitrarily weak Zeeman fields.
Second, in the deep subgap case, we have determined the conditions
for observing a pair of topologically unprotected yet spatially 
separated Majorana bound states in a double dot.  The formation
of such exotic particles is presently under vigorous study and could
be indirectly detected in the CPR through the
critical phases $\varphi$, where the current switches from a finite value
to zero.  We hope that these effects can soon be observed
in experiments.

\acknowledgments
We thank Roland H\"utzen for help in preparing the figures. This work
has been supported by the DFG (Grant No. EG 96/9-1).


\begin{thebibliography}{12}
\bibitem{footcot}
Perturbation theory may still fail when the excitation energy
$\epsilon_b$ entering Eq.~\eqref{Q123} is close to zero (i.e., smaller
than the hybridization scales). For instance, in Eq.~\eqref{Q123} with 
$i=1$, this causes a divergence of the supercurrents
$I_{a,0} \propto (\Gamma / \Delta)^2 (\Delta / \epsilon_b) \to \infty$.
In such a case, the addition (or removal) of 
a single Cooper pair to (from) the dot leaves
the ground-state energy unchanged. This may happen 
when the single-particle energy differences are
much smaller than $\Delta$ and we have large $E_c$ with $N_0 =n_g \pm 1$.
\bibitem{foot4}
The double-frequency integrals in Eq.~\eqref{Q123} 
can be carried out analytically by replacing $f(\tau)\to\tilde f(\tau)$
in Eq.~\eqref{fdef}, with $(\omega^2+\Delta^2)^{-1/2}\to (\tilde \Delta^2/
\Delta)/(\omega^2+\tilde\Delta^2)$, 
where $\tilde \Delta$ is a fitting
parameter of order $\Delta$ chosen such that $f(\tau)$ and $\tilde f(\tau)$ 
have the same low-energy behavior. 
Within $1\%$ accuracy, analytical results (using $\tilde f$) were found to  
match numerics (using  $f$) for all parameters studied. 
We have used $f\to\tilde f$ exclusively for 
generating the figures in Sec.~\ref{sec4}.
\bibitem{footnote1}
Note that for $N_0=1$, the ground-state is two-fold spin degenerate.
Since both spin orientations lead to the same result, our derivation 
(assuming non-degenerate $|G\rangle$) still holds.
\bibitem{foot5}
Note that for $\epsilon \to 0$, Eq.~\eqref{calI23chi} 
coincides with Eq.~\eqref{calI12eps}.
\end{thebibliography}
\end{document}